\documentclass[conference]{IEEEtran}
\IEEEoverridecommandlockouts
\usepackage{amsmath,amsfonts}
\usepackage{amssymb}
\usepackage{algorithm}
\usepackage{algpseudocode}
\algrenewcommand\algorithmicrequire{\textbf{Input:}}
\algrenewcommand\algorithmicensure{\textbf{Output:}}
\usepackage{array}
\usepackage[caption=false,font=footnotesize,labelfont=rm,textfont=rm]{subfig}
\usepackage{textcomp}
\usepackage{stfloats}
\usepackage{url}
\usepackage{verbatim}
\usepackage{graphicx}
\usepackage{cite}
\def\BibTeX{{\rm B\kern-.05em{\sc i\kern-.025em b}\kern-.08em
    T\kern-.1667em\lower.7ex\hbox{E}\kern-.125emX}}
\begin{document}

\title{Analysis and Compensation of Tx and Rx IQ\\ Imbalances in AFDM System
{}
}

\author{
\IEEEauthorblockN{Hongjun~Liu$^{\ast}$, Liaoyuan~Zeng$^{\ast}$, Junhao~Tian$^{\ast}$, Qingyu~Li$^{\ast}$, Fuchen~Xu$^{\ast}$, Chengxiang~Liu$^{\ast}$, Guanghui~Liu$^{\ast}$}
\IEEEauthorblockA{{$^{\ast}$School of Information and Communication Engineering, University of Electronic Science and Technology of China,}\\
Chengdu, China \\
Email: \{hjliu@std., Lyzeng@, 202444060140@std., qingyu.li@std., fuchenxu@std., cxliu@std., guanghuiliu@,\}uestc.edu.cn}
}

\maketitle

\begin{abstract}
Affine frequency division multiplexing (AFDM) is a recently proposed multicarrier waveform whose bit error rate (BER) performance in doubly selective channels is comparable to that of orthogonal time-frequency space (OTFS) and superior to that of orthogonal frequency division multiplexing (OFDM). In this paper, the impacts of joint transmitter (Tx) and receiver (Rx) in-phase and quadrature imbalance (IQI) on AFDM signals are investigated, where we show that AFDM suffers more severe IQI than OFDM and OTFS due to the inherent feature of complicated chirp-assisted modulation. We further derive analytical expressions for the pairwise and average bit error probability as a function of the IQI parameters. These indicate that such distortions significantly limit the achievable operating signal-to-noise ratio at the receiver side and data rates. To this end, we propose a cascade compensation scheme to mitigate these effects. Specifically, we first compensate for Rx IQI to convert the improper Gaussian noise into additive white Gaussian noise, and then apply a judicious design to eliminate the Tx IQI. Both analytical and simulation results reveal that joint Tx and Rx IQI introduce an error floor in the BER performance of AFDM systems, whereas the proposed approach effectively compensates such impairments. 
\end{abstract}


\section{Introduction}
Next-generation mobile communication systems are anticipated to support a wide range of data services including ultra-reliable, high-rate, and low-latency communications in highly dynamic environments \cite{9040264}, such as high-speed railway, unmanned aerial vehicle (UAV), and satellite-to-ground communications. The channels therein are typically characterized by highly time-varying and severe delay-Doppler dispersions, which pose significant challenges to the current widely deployed orthogonal frequency division multiplexing (OFDM) waveform \cite{9508932}. Extensive efforts have therefore been devoted to developing new waveforms which are capable of operating effectively over such challenging channels. Orthogonal time-frequency space (OTFS), as a prominent solution, employs a two-dimensional modulation framework in which the information symbols are multiplexed in the delay-Doppler domain and thus a quasi-static channel representation is yielded \cite{10891132}. Although OTFS outperforms OFDM over such linear time-varying channels, it suffers from large pilot overhead. Another promising candidate is affine frequency division multiplexing (AFDM) \cite{9746329}, a recently proposed waveform in which information symbols are carried by a set of orthogonal chirps defined in the discrete affine Fourier transform (DAFT) domain and mapped into the time domain through the inverse DAFT (IDAFT). In particular, by adaptively tuning the chirp rate according to the Doppler profile, AFDM enables a separable and quasi-static channel representation, therefore achieving full diversity in doubly selective channels (DSC). The bit error rate (BER) performance of AFDM is demonstrated in \cite{10087310}, where a minimum mean-square error (MMSE) detector is introduced.

Direct-conversion radio-frequency (RF) transceivers \cite{482187}, owing to their potential for low-cost and low-power implementation in silicon, have been widely adopted in modern communication systems. However, such architectures inevitably induce in-phase and quadrature-phase imbalances (IQI) for communication signals. The IQI can severely limit the achievable operating signal-to-noise ratio (SNR) at the receiver and, consequently, the supported constellation size and data rates \cite{1468518}. Furthermore, we investigate that the implementation of AFDM-based physical layers is more susceptible to IQI induced distortion in the front-end analog processing than OFDM and OTFS counterparts. This is mainly because the more sophisticated chirp-assisted multicarrier modulation of AFDM gives rise to more severe mirror interference. More critically, the joint transmitter (Tx) and receiver (Rx) IQI often exist in practical systems, where conventional receiver schemes are completely no longer sufficient.
\addtolength{\topmargin}{-0.18in}
For such systems impaired by joint Tx and Rx IQI, widely linear processing paired with IQI compensation is generally recognized as an effective solution and has been extensively applied to OFDM \cite{5710591,4939424} and OTFS\cite{9714472,9543616}. Similarly, a widely linear MMSE (WL-MMSE) detector is proposed in Rx IQI effected AFDM system \cite{2025arXiv251210036}, where the covariance and pseudo-covariance of improper Gaussian noise are exploited. Specifically, the additive white Gaussian noise (AWGN) in the original system is transformed into improper Gaussian noise owing to Rx IQI. However, widely linear processing owing to jointly process the signal and the counterpart of conjugate incurs high computational complexity, resulting in substantial computational overhead.

In this paper, we first derive the input-output relation of AFDM under joint Tx and Rx IQI and further obtain analytical expressions for the pairwise error probability (PEP) as well as the average bit error probability (ABEP) in terms of the IQI parameters. To suppress the resulting distortion, we propose a cascaded compensation scheme that can be integrated with any detector with negligible additional computational overhead. Both analytical and simulation results show that Tx and Rx IQI cause an error floor in the BER performance of AFDM systems. Nevertheless, the proposed scheme can effectively compensate for these impairments.

\section{System Model}
This section reviews the basic AFDM signal model and then introduces the IQI distortion in such system, where the performance analysis is given.

\subsection{AFDM Transceiver With Ideal RF Front-end}
Considering an AFDM symbol consisting of $N$ chirp-subcarriers, the IDAFT is performed to map the information symbol $x[m]$ in the DAFT domain to the time domain signal $s[n]$ as
\begin{equation}
    \label{eq1}
    s[n]=\frac1{\sqrt{N}}\sum_{m=0}^{N-1}x[m]e^{j2\pi(c_1n^2+c_2m^2+nm/N)},
\end{equation}
\noindent where $c_1$ as well as $c_2$ are AFDM parameters. Besides, the matrix form of (1) is given by $\mathbf{s}=\mathbf{\Lambda}_{c_1}^H\mathbf{F}^H\mathbf{\Lambda}_{c_2}^H\mathbf{x}$, of which $\mathbf{\Lambda}_{c}$ denote a diagonal matrix with entries $e^{-j2\pi cn^{2}}$ and $\mathbf{F}$ implies discrete Fourier transform (DFT) matrix. Before transmitting $\mathbf{s}$, an chirp-periodic prefix (CPP) is added.

The channel impulse response of the DSC is characterized as
\begin{equation}
    \label{eq2}
    h(\tau,\nu)=\sum_{i=1}^{P}h_{i}\delta(\tau-\tau_i)\delta(\nu-\nu_{i})\:,
\end{equation}
\noindent where $h_i$, $\tau_i$ and $\nu_i$ are the channel coefficient, delay and Doppler shift associated with the $i$-th path. Meanwhile, $P$ denotes the number of paths, and the supportive bound of $\tau_i$ and $\nu_i$ are $\left[0, \tau_\text{max}\right]$ and $\left[-\nu_\text{max},\nu_\text{max}\right]$, where $\tau_\text{max}$ and $\nu_\text{max}$ denote the maximum delay spread and Doppler shift. In this paper, the $\tau_\text{max}$ and $\nu_\text{max}$ are assumed to be known in advance.

After discarding CPP, the received signal $r[n]$ in the time domain is expressed as
\begin{equation}
    \label{eq3}
    r[n]=\sum_{i=1}^Ph_ie^{-j\frac{2\pi}{N} \nu_in}s[(n-\tau_i)_N]+w[n] ,
\end{equation}
\noindent where $w[n]{\sim}\mathcal{CN}(0,\sigma^2)$ represents additive white Gaussian noise (AWGN). Followed by AFDM demodulation, the received signal in the DAFT domain is written as
\begin{equation}
    \label{eq4}
    y[m]=\frac1{\sqrt{N}}\sum_{n=0}^{N-1}r[n]e^{-j2\pi(c_1n^2+c_2m^2+nm/N)}.
\end{equation}

Hence, the output in the matrix representation is formulated as
\begin{equation}
    \label{eq5}
    \mathbf{y}=\sum_{i=1}^{P}h_{i}\mathbf{A}\boldsymbol{\Gamma}_{\mathrm{CPP}_{i}}\boldsymbol{\Delta}_{\nu_{i}}\boldsymbol{\Pi}^{\tau_{i}}\mathbf{A}^H\mathbf{x}+\tilde{\mathbf{w}}=\mathbf{A}\mathbf{H}\mathbf{A}^H\mathbf{x}+\tilde{\mathbf{w}} ,
\end{equation}

\noindent where $\mathbf{H}=\sum_{i=1}^{P}h_{i}\boldsymbol{\Gamma}_{\mathrm{CPP}_{i}}\boldsymbol{\Delta}_{\nu_{i}}\boldsymbol{\Pi}^{\tau_{i}}$ denotes effective channel impulse in the time domain, $\boldsymbol{\Pi}$ models as forward cyclic-shift matrix and $\boldsymbol{\Delta}$ is Dopper shift matrix. Additionally, $\mathbf{A}=\mathbf{\Lambda}_{c_2}\mathbf{F}\mathbf{\Lambda}_{c_1}$ is a unitary matrix, and $\tilde{\mathbf{w}}=\mathbf{A}\mathbf{w}$ denotes the Gaussian noise in the DAFT domain. Besides, to achieve the optimal diversity order in AFDM systems, the $c_1=\frac{2\left(\nu_\text{max}+\zeta_\nu\right)+1}{2N}$ and $c_2$ set to a rational number sufficiently smaller than $\frac{1}{2N}$ are considered \cite{10087310}, where $\zeta_\nu\geq0$ is used to combat the fractional Doppler. 

\subsection{IQ Imbalance}
The hardware impairments are not introduced in AFDM signal under previous discussion, whereas IQI are inevitably occurred in practical direct-conversion RF passed signal. Considering the continuous-time domain signals $s(t)$, $r(t)$, and the narrowband IQI model the output signal of Tx is given by
\begin{equation}
    \label{eq6}
    \bar{s}(t)=\mu_\text{tx}s(t)+\upsilon_\text{tx}s^*(t),
\end{equation}
\noindent where $\mu_\text{tx}=\cos(\theta_\text{tx}/2)+j\alpha_\text{tx}\sin(\theta_\text{tx}/2)$ and $\upsilon_\text{tx}=\alpha_\text{tx}\cos(\theta_\text{tx}/2)-j\sin(\theta_\text{tx}/2)$ are distortion parameters, of which $\theta_\text{tx}$ and $\alpha_\text{tx}$ represent the phase and the amplitude mismatches between I and Q branches \cite{1468518}. Moreover, the Tx amplitude imbalance, stated in dB, is defined as $\mathrm{AIm}_\text{tx}=10\mathrm{log}(1+\alpha_\text{tx})$, and the Tx phase imbalance denotes $\mathrm{PIm}_\text{tx}=\theta_\text{tx}$. Furthermore, a similar expression is used to model IQI at the Rx, derived as
\begin{equation}
    \label{eq7}
    \bar{r}(t)=\mu_\text{rx}r(t)+\upsilon_\text{rx}r^{*}(t)\:,
\end{equation}
\noindent where $\mu_\text{rx}$ and $\upsilon_\text{rx}$ are the corresponding Rx IQI parameters and are defined analogously to their Tx counterparts, $\mu_\text{tx}$ and $\upsilon_\text{tx}$.

Assume that the Nyquist criterion is satisfied, and \eqref{eq5} is rewritten as 
\begin{align}
    \label{eq8}
    \mathbf{y}=\mathbf{A}\mu_\text{rx}&\left(\mu_\text{tx}\mathbf{H}\mathbf{A}^H\mathbf{x}+     \upsilon_\text{tx}\mathbf{H}\mathbf{A}^T\mathbf{x}^*\right) \nonumber \\
    &+\mathbf{A}\upsilon_\text{rx}\left(\mu_\text{tx}^*\mathbf{H}^*\mathbf{A}^T\mathbf{x}^*+\upsilon_\text{tx}^*\mathbf{H}^*\mathbf{A}^H\mathbf{x}\right)+\bar{\mathbf{w}}
\end{align}
\noindent to present the matrix form of the DAFT domain received signal distorted by joint Tx and Rx IQI. In addition, $\bar{\mathbf{w}}=\mathbf{A}\left(\mu_\text{rx}\mathbf{w}+\upsilon_\text{rx}\mathbf{w}^*\right)$ denote the improper Gaussian noise in the DAFT domain.
\addtolength{\topmargin}{-0.1in}
Furthermore, we evaluate the performance of AFDM system by considering joint Tx and Rx IQI with different channels, shown in Fig. 1. The performance of OFDM and OTFS are considered as the benchmarks, where the number of subcarriers is set to 256, and the number of symbols is 16. For fair comparison, the simulation builds upon quadrature phase shift keying (QPSK) modulation with MMSE detector for all waveforms, and IQI setting as $\mathrm{AIm}_\text{tx}=\mathrm{AIm}_\text{rx}=1.5$ dB, $\mathrm{PIm}_\text{tx}=\mathrm{PIm}_\text{rx}=3.5^\circ$. At a target BER of $10^{-3}$ in Fig. 1, compared to the ideal system with no IQI, AFDM undergoes an SNR loss of 7.2 dB in AWGN channel, whereas OFDM and OTFS experience an SNR loss of 3 dB. Likewise, at a target BER of $1.9\times10^{-2}$, AFDM undergoes an SNR loss of 3 dB in DSC, while the losses are 1 dB and 2.5 dB for OTFS and OFDM system respectively. The reason that AFDM suffers more severe losses is the complex-conjugate operations in the DAFT domain.

\begin{figure}[!t]
\label{fig1}
\centering
\subfloat[AWGN channel.]{
		\includegraphics[scale=0.55]{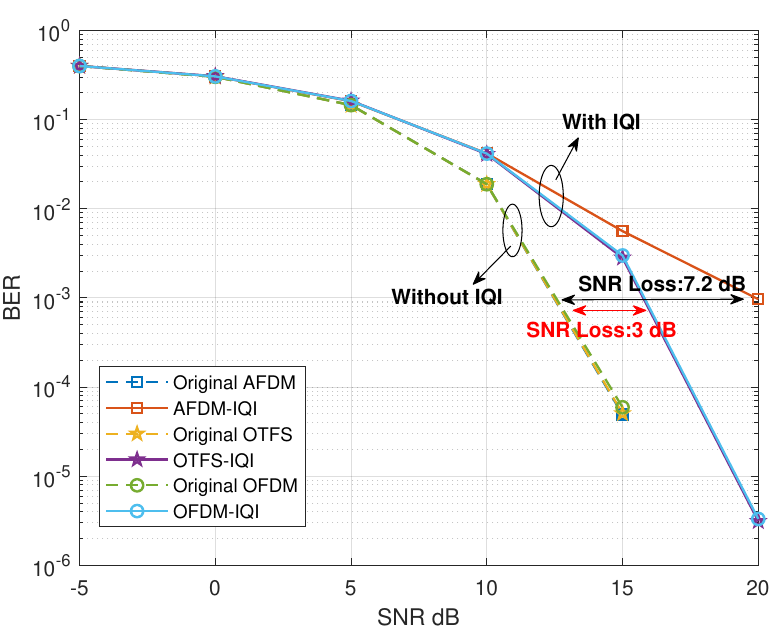}}
        
\subfloat[Time-varying channel.]{
		\includegraphics[scale=0.55]{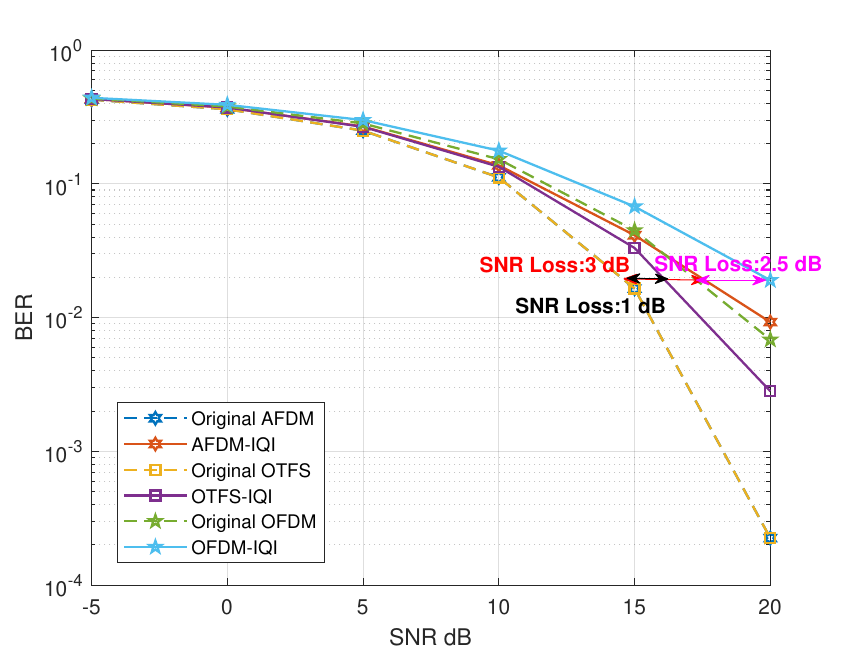}}
\caption{BER performance of QPSK modulated AFDM with MMSE detector compared to OFDM and OTFS.}
\end{figure}

\subsection{The Input-Output Relations of IQ Imbalanced AFDM System}
\begin{figure}[!t]
\centering
\includegraphics[width=3.7in]{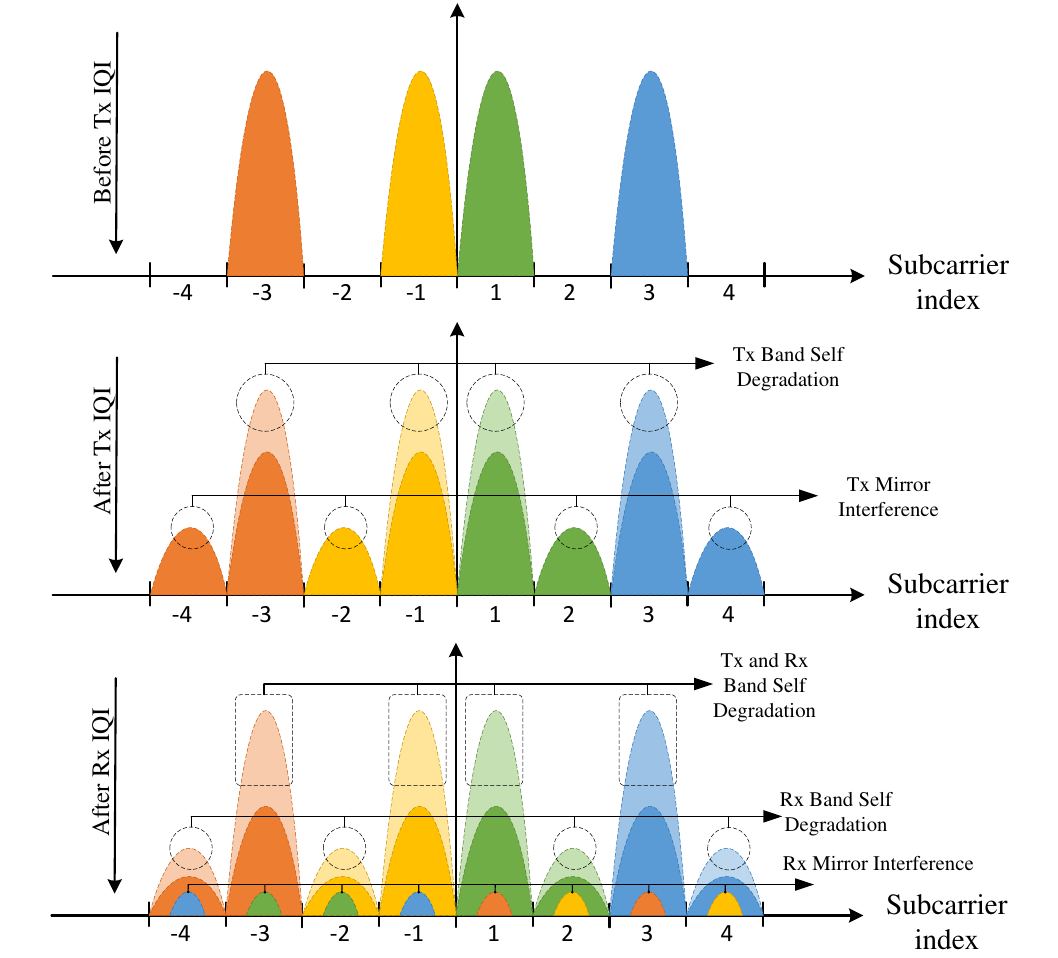}
\caption{Spectra illustration of the joint effect of TX and Rx IQ imbalance for AFDM signals.}
\label{fig2}
\end{figure}

To gain insight into the origin of these impairments, we derive the system input–output relationship from the perspective of chirp subcarriers, which can be expressed as 
\begin{align}
    \label{eq9}
    \mathbf{y}&=\underbrace{\mu_\text{rx}\mathbf{AHA}^{H}\mu_\text{tx}\mathbf{x}}_{\text{attenuated transmitted}}+\underbrace{\mu_\text{rx}\mathbf{AHA}^{T}\upsilon_\text{tx}\mathbf{x}^{*}}_{\text{mirror interference}} \nonumber \\
    &+\underbrace{\upsilon_\text{rx}\mathbf{AH}^{*}\mathbf{A}^{T}\mu_\text{tx}^{*}\mathbf{x}^{*}+\upsilon_\text{rx}\mathbf{AH}^{*}\mathbf{A}^{H}\upsilon_\text{tx}^{*}\mathbf{x}}_{\text{Tx IQI affected mirror-chunk interference}}+\underbrace{\mathbf{\bar{w}}}_{\text{amplified}}.
\end{align}
\noindent It is evident from the resulting expression that the signal is not only attenuated during transmission but also corrupted by amplified noise. Moreover, the interference arises not only from mirror interference of Rx IQI but also from the mirror-chunk interference induced by Tx IQI. To elucidate this process, a graphical illustration is provided in Fig. \ref{fig2}. For an ideal AFDM signal, the presence of Tx IQI attenuates the desired signal and induces an initial image component. Upon reception, the signal undergoes further degradation due to Rx IQI. This not only causes additional attenuation to the desired component and distorts the image component generated by Tx IQI, but also generates a new image component. Consequently, this cascaded distortion inevitably triggers inter-carrier interference.

\section{Error Performance Analysis}
In this section, thanks to the system input-output relation obtained in \eqref{eq9} in the DAFT domain and by assuming the optimal maximum likelihood (ML) criterion, the ABEP can be obtained in the presence of joint Tx and Rx IQI distortion. 

For further derivation, \eqref{eq9} can be rewritten as
\begin{equation}
    \label{eq10}
    \mathbf{y}=\boldsymbol{\Psi}_1(\mathbf{x})\mathbf{h}+\boldsymbol{\Psi}_2(\mathbf{x})\mathbf{h}^*+\bar{\mathbf{w}}\
\end{equation}
\noindent where for $i=1,2,\dots,P$ we have
\begin{equation}
    \label{eq11}
    \boldsymbol{\Psi}_1(\mathbf{x})=\left[\boldsymbol{\Psi}_{1,1}(\mathbf{x}),\boldsymbol{\Psi}_{1,2}(\mathbf{x}),\dots,\boldsymbol{\Psi}_{1,P}(\mathbf{x})\right]\in\mathbb{C}^{N\times P},
\end{equation}
\begin{equation}
    \label{eq12}
    \boldsymbol{\Psi}_2(\mathbf{x})=\left[\boldsymbol{\Psi}_{2,1}(\mathbf{x}),\boldsymbol{\Psi}_{2,2}(\mathbf{x}),\dots,\boldsymbol{\Psi}_{2,P}(\mathbf{x})\right]\in\mathbb{C}^{N\times P},
\end{equation}
\begin{align}
    \label{eq13}
    \boldsymbol{\Psi}_{1,i}(\mathbf{x})=\mu_\text{rx}\mu_\text{tx}\mathbf{A}&\boldsymbol{\Gamma}_{\mathrm{CPP}_{i}}\boldsymbol{\Delta}_{\nu_{i}}\boldsymbol{\Pi}^{\tau_{i}}\mathbf{A}^H\mathbf{x} \nonumber \\
    &+\mu_\text{rx}\upsilon_\text{tx}\mathbf{A}\boldsymbol{\Gamma}_{\mathrm{CPP}_{i}}\boldsymbol{\Delta}_{\nu_{i}}\boldsymbol{\Pi}^{\tau_{i}}\mathbf{A}^T\mathbf{x}^*,
\end{align}
\begin{align}
    \label{eq14}
    \boldsymbol{\Psi}_{2,i}(\mathbf{x})=\upsilon_\text{rx}\mu_\text{tx}^*\mathbf{A}&\left(\boldsymbol{\Gamma}_{\mathrm{CPP}_{i}}\boldsymbol{\Delta}_{\nu_{i}}\boldsymbol{\Pi}^{\tau_{i}}\right)^*\mathbf{A}^T\mathbf{x}^* \nonumber \\
    &+\upsilon_\text{rx}\upsilon_\text{tx}^*\mathbf{A}\left(\boldsymbol{\Gamma}_{\mathrm{CPP}_{i}}\boldsymbol{\Delta}_{\nu_{i}}\boldsymbol{\Pi}^{\tau_{i}}\right)^*\mathbf{A}^H\mathbf{x},
\end{align}
\begin{equation}
    \label{eq15}
    \mathbf{h}=\left[h_1,h_2,\dots,h_P\right]\in\mathbb{C}^{P\times1}.
\end{equation}
\noindent Apparently, $\bar{\mathbf{w}}$ is a linear combination of Gaussian vectors $\mathbf{w}$ and $\mathbf{w}^*$. Therefore, $\bar{\mathbf{w}}$ stays Gaussian and since $\mathbf{w}$ is zero-mean, $\bar{\mathbf{w}}$ becomes zero-mean. Also, by considering the unitary property of DAFT, the covariance matrix of $\bar{\mathbf{w}}$ is given as
\begin{align}
    \label{eq16}
    \mathrm{cov}(\bar{\mathbf{w}})&=\mathbb{E}(\bar{\mathbf{w}}\bar{\mathbf{w}}^H) \nonumber \\
    &=\mathbf{A}\{|\mu_\text{rx}|^2\mathbb{E}(\mathbf{w}\mathbf{w}^H)+\mu_\text{rx}\upsilon_\text{rx}\mathbb{E}(\mathbf{w}\mathbf{w}^T) \nonumber \\
    &+\mu_\text{rx}\upsilon_\text{rx}\mathbb{E}(\mathbf{w}^*\mathbf{w}^H)+|\upsilon_\text{rx}|^2\mathbb{E}(\mathbf{w}^*\mathbf{w}^T)\}\mathbf{A}^H \nonumber \\
    &=\left(|\mu_\text{rx}|^2+|\upsilon_\text{rx}|^2\right)\sigma^2\mathbf{I}_N,
\end{align}
\noindent where $\mathbb{E}(\mathbf{w}\mathbf{w}^T)=\mathbb{E}(\mathbf{w}^*\mathbf{w}^H)=\mathbf{0}$. Moreover, the power disparities and cross-correlation between the in-phase (I) and quadrature (Q) branches caused by hardware IQ impairments break the circular symmetry of ideal white noise, thereby inevitably resulting in a nonzero pseudo-covariance under the widely statistical model. Thus, the pseudo-covariance of $\bar{\mathbf{w}}$ is formulated as
\begin{align}
    \label{eq17}
    \mathrm{pcov}(\bar{\mathbf{w}})&=\mathbb{E}(\bar{\mathbf{w}}\bar{\mathbf{w}}^T) \nonumber \\
    &=\mathbf{A}\{|\mu_\text{rx}|^2\mathbb{E}(\mathbf{w}\mathbf{w}^T)+\mu_\text{rx}\upsilon_\text{rx}\mathbb{E}(\mathbf{w}\mathbf{w}^H) \nonumber \\
    &+\mu_\text{rx}\upsilon_\text{rx}\mathbb{E}(\mathbf{w}^*\mathbf{w}^T)+|\upsilon_\text{rx}|^2\mathbb{E}(\mathbf{w}^*\mathbf{w}^H)\}\mathbf{A}^T \nonumber \\
    &=2\mu_\text{rx}\upsilon_\text{rx}\sigma^2\mathbf{A}\mathbf{A}^T\:,
\end{align}
\noindent where $\mathbb{E}(\mathbf{w}^*\mathbf{w}^T)=\mathbb{E}\left((\mathbf{w}\mathbf{w}^H)^*\right)=\sigma^2\mathbf{I}_N$. Thus, the Rx IQI induced improper noise $\bar{\mathbf{w}}$ in the DAFT domain is distributed as $\mathcal{CN}\left(\mathbf{0},\left(|\mu_\text{rx}|^2+|\upsilon_\text{rx}|^2\right)\sigma^2\mathbf{I}_N,2\mu_\text{rx}\upsilon_\text{rx}\sigma^2\mathbf{A}\mathbf{A}^T\right)$, where $\sigma^2_{\bar{\mathbf{w}}}$ is equaled to$\left(|\mu_\text{rx}|^2+|\upsilon_\text{rx}|^2\right)\sigma^2$. 
\addtolength{\topmargin}{-0.1in}
Accordingly, the ML criterion is expressed as
\begin{equation}
    \label{eq18}
    \hat{\mathbf{x}}_{\mathrm{ML}}=\min_{\mathbf{x}\in\mathbb{C}^{N\times1}}\|\mathbf{y}-\boldsymbol{\Psi}_1(\mathbf{x})\mathbf{h}-\boldsymbol{\Psi}_2(\mathbf{x})\mathbf{h}^*\|^2.
\end{equation}
\noindent To begin, we derive a tight upper bound for conditional PEP. Given the perfect CSI vector $\mathbf{h}$, the PEP is defined as the probability of erroneously deciding in favor of symbol $x[q]$ when $x[p]$ is actually transmitted. By defining the equivalent codeword matrix as $\boldsymbol{\Psi}(\mathbf{x})=\boldsymbol{\Psi}_1(\mathbf{x})+\boldsymbol{\Psi}_2(\mathbf{x})$, the conditional PEP bound for ML detection under Tx and Rx IQI can be written as
\begin{equation}
    \label{eq19}
\Pr(x[p]\to x[q]|\mathbf{h})\leq\mathbb{Q}\left(\sqrt{\frac{\left\|\left(\boldsymbol{\Psi}(x[p])-\boldsymbol{\boldsymbol{\Psi}}(x[q])\right)\mathbf{h}\right\|^2}{2\sigma_\mathbf{\bar{w}}^2}}\right).
\end{equation}
Direct evaluation of \eqref{eq19} is mathematically cumbersome in practice. To circumvent this intractability and obtain an analytical expression, we employ a highly accurate exponential approximation for the Gaussian Q-function, namely $\mathbb{Q}(x)\leq e^{-x^2/2}/12+e^{-2x^2/3}/4$. By taking the expectation with respect to $\mathbf{h}$, the unconditional PEP is bounded by
\begin{align}
    \label{eq20}
    &\Pr(x[p]\to x[q])\leq \nonumber \\
    &\mathbb{E}\left(\frac{1}{12}e^{-\gamma_1\|\left(\boldsymbol{\Psi}(x[p])-\boldsymbol{\Psi}(x[q])\right)\mathbf{h}\|^2}+\frac{1}{4}e^{-\gamma_2\|\left(\boldsymbol{\Psi}(x[p])-\boldsymbol{\Psi}(x[q])\right)\mathbf{h}\|^2}\right),
\end{align}
\noindent where the dimension parameter satisfying $\gamma_1=\frac{1}{4\sigma^2_{\bar{\mathbf{w}}}}$ and $\gamma_2=\frac{1}{3\sigma^2_{\bar{\mathbf{w}}}}$. Besides, the probability density function of Gaussian random vector $\mathbf{h}$ is characterized by
\begin{equation}
    \label{eq21}
    f(\mathbf{h})=\frac{\pi^{-\epsilon}}{\mathrm{det}\left(\mathbb{E}(\mathbf{h}\mathbf{h}^H)\right)}e^{-\mathbf{h}^H\left(\mathbb{E}(\mathbf{h}\mathbf{h}^H)\right)^{-1}\mathbf{h}},
\end{equation}
\noindent where $1\leq\epsilon=\mathrm{rank}(\mathbb{E}(\mathbf{h}\mathbf{h}^H))\leq P$. Substituting \eqref{eq21} into the expectation calculation of \eqref{eq20} and using the spectral theorem, the final PEP upper bound is derived as
\begin{equation}
    \label{eq22}
    \Pr(x[p]\to x[q])\leq\frac{1}{12}\prod_{k=1}^{\mathcal{K}}\frac{1}{1+\frac{\gamma_{1}\eta_{k}}{P}}+\frac{1}{4}\prod_{k=1}^{\mathcal{K}}\frac{1}{1+\frac{\gamma_{2}\eta_{k}}{P}},
\end{equation}
\noindent where $\eta_{k}$ denote the non-zero eigen values of $\left(\boldsymbol{\Psi}(x[p])-\boldsymbol{\Psi}(x[q])\right)^H\left(\boldsymbol{\Psi}(x[p])-\boldsymbol{\Psi}(x[q])\right)$, and $\mathcal{K}=\mathrm{rank}\left(\mathbb{E}(\mathbf{h}\mathbf{h}^H)\left(\boldsymbol{\Psi}(x[p])-\boldsymbol{\Psi}(x[q])\right)^H\left(\boldsymbol{\Psi}(x[p])-\boldsymbol{\Psi}(x[q])\right)\right)$.

\addtolength{\topmargin}{-0.1in}

Ultimately, the ABEP performance of the AFDM system in the presence of transceiver IQI can be evaluated through the union bounding approach. By incorporating \eqref{eq22}, this theoretical upper bound is formulated as
\begin{equation}
    \label{eq23}
    \mathrm{ABEP}\leq \frac{1}{N_b2^{N_b}}\sum_{x[p]}\sum_{x[q]}\Pr(x[p]\to x[q])N_{be}(x[p],x[q]),
\end{equation}
\noindent where $N_b$ represents the bit length and $N_{be}(x[p],x[q])$ denotes the pairwise bit error distance, i.e., the number of erroneous bits incurred when $x[p]$ is erroneously detected as $x[q]$.

\section{Proposed Cascaded Compensation Approach}
In the preceding section, we theoretically analyzed the system performance under optimal detection in the absence of IQI compensation. Specifically, a theoretical upper bound for the ABEP is derived as a function of the IQI parameters. Furthermore, to effectively mitigate the IQI distortion, this section proposes a cascaded compensation scheme, which is compatible with various AFDM detection methods with only linear additional complexity. This approach first performs Rx IQI compensation in the time domain to transform the improper Gaussian noise back into AWGN, and followed by Tx IQI compensation applied to the post-detection symbols.

\subsection{Rx IQI Compensation}
Compensating for IQI is an imperative step, given that the adverse effects must be mitigated before data symbols can be reliably detected. Because the Tx IQI is subsequently mirrored by Rx IQI, the Rx IQI compensation is executed first. For this purpose, a low-complexity scheme is employed rather than conventional linear processing methods. Considering that IQI inherently manifests in the time domain, we propose a time-domain IQI compensation method, denoted as
\begin{align}
    \label{eq24}
    \mathbf{r}_\text{comp}=\frac{\mu_\text{rx}^*\bar{\mathbf{r}}-\upsilon_\text{rx}\bar{\mathbf{r}}^*}{|\mu_\text{rx}|^2-|\upsilon_\text{rx}|^2},
\end{align}
\noindent where the improper noise is completely eliminated following this compensation procedure, expressed as
\begin{align}
    \label{eq25}
    \mathbf{w}_\text{comp}=&\frac{\mu_\text{rx}^*\left(\mu_\text{rx}\mathbf{w}+\upsilon_\text{rx}\mathbf{w}^*\right)-\upsilon_\text{rx}\left(\mu_\text{rx}\mathbf{w}+\upsilon_\text{rx}\mathbf{w}^*\right)^*}{|\mu_\text{rx}|^2-|\upsilon_\text{rx}|^2} \nonumber \\
    =&\frac{|\mu_\text{rx}|^2\mathbf{w}-|\upsilon_\text{rx}|^2\mathbf{w}}{|\mu_\text{rx}|^2-|\upsilon_\text{rx}|^2} \nonumber \\
    =&\mathbf{w}.
\end{align}
Subsequently, the time domain received signal, now free from Rx IQI, is mapped into the DAFT domain through 
\begin{align}
    \label{eq26}
    \mathbf{y}_\text{comp}&=\mathbf{A}\mathbf{r}_\text{comp} \nonumber \\
    &=\mathbf{A}\mathbf{H}(\mu_\text{tx}\mathbf{A}^H\mathbf{x}+\upsilon_\text{tx}\mathbf{A}^T\mathbf{x}^*)+\mathbf{A}\mathbf{w}_\text{comp}.
\end{align}
\subsection{Tx IQI Compensation}
To facilitate the further derivations, \eqref{eq26} can be equivalently rewritten as
\begin{equation}
    \label{eq27}
    \mathbf{y}_\text{comp}=\mathbf{A}\mathbf{H}\mathbf{A}^H\mathbf{A}(\mu_\text{tx}\mathbf{A}^H\mathbf{x}+\upsilon_\text{tx}\mathbf{A}^T\mathbf{x}^*)+\mathbf{A}\mathbf{w}_\text{comp},
\end{equation}
\noindent by leveraging the unitary property $\mathbf{A}^H\mathbf{A}=\mathbf{I}$. Remarkably, the aforementioned equation can be simplified as
\begin{equation}
    \label{eq28}
    \mathbf{y}_\text{comp}=\mathbf{H}_\text{eff}\tilde{\mathbf{x}}+\mathbf{A}\mathbf{w}_\text{comp},
\end{equation}
\noindent where $\tilde{\mathbf{x}}=\mu_\text{tx}\mathbf{x}+\upsilon_\text{tx}\mathbf{A}^T\mathbf{A}\mathbf{x}^*$ and $\mathbf{H}_\text{eff}=\mathbf{A}\mathbf{H}\mathbf{A}^H$. Clearly, under this condition, where the equivalent channel response $\mathbf{H}_\text{eff}$ is proper as well as sparse and the noise is distributed as proper complex Gaussian, an arbitrary AFDM detector can be employed to perform the coarse estimation of the data symbols. Herein, we adopt the MMSE detector, whose closed-form expression is given by
\begin{equation}
    \label{eq29}
    \hat{\tilde{\mathbf{x}}}=\mathbf{H}_\text{eff}^H\left(\mathbf{H}_\text{eff}\mathbf{H}_\text{eff}^H+\sigma^2\mathbf{I}\right)^{-1}\mathbf{y}_\text{comp}.
\end{equation}
\noindent After acquiring the coarse data symbols $\hat{\tilde{\mathbf{x}}}$, the following low-complexity Tx IQI compensation scheme is applied
\begin{equation}
    \label{eq30}
    \hat{\mathbf{x}}=\mathbf{A}\frac{\mu_\text{tx}^*\mathbf{A}^H\hat{\tilde{\mathbf{x}}}-\upsilon_\text{tx}\mathbf{A}^T\hat{\tilde{\mathbf{x}}}}{|\mu_\text{tx}|^2-|\upsilon_\text{tx}|^2}.
\end{equation}
\noindent At this point, the AFDM signals impaired by joint Tx and Rx IQI have been fully compensated.
\subsection{Complexity Analysis}
By completely circumventing high-dimensional matrix operations, the proposed compensation approach achieves exceedingly low computational complexity. Furthermore, by restoring the sparse equivalent channel and the proper complex Gaussian noise in the DAFT domain, it can be seamlessly integrated with any arbitrary AFDM detection algorithm. The proposed cascaded compensation scheme primarily consists of the operations in \eqref{eq24} and \eqref{eq30}, where each operation requires $\mathcal{O}(N)$, resulting in an overall computational complexity of $\mathcal{O}(2N)$. When paired with popular AFDM detectors, such as MMSE, maximal ratio combining, and expectation propagation, which are high complexity far exceeding $\mathcal{O}(2N)$, the additional computational overheads of the proposed compensated method are marginal and can be negligible.

\section{Simulation Results}
In this section, we evaluate the performance of the AFDM system under the influence of joint Tx and Rx IQI. Monte Carlo simulations are conducted to validate the derived analytical ABEP bounds and to verify the effectiveness of the proposed compensation scheme. We consider a direct-conversion RF transceiver operating at a high mobility speed of 540 km/h, employing QPSK modulation with $N=64$ chirp-subcarriers. A comprehensive list of the simulation parameters is provided in Table \uppercase\expandafter{\romannumeral1}. Furthermore, it is assumed throughout the following evaluations that perfect CSI is available at the receiver.

\begin{table}[t]
\caption{\textbf{SIMULATION PARAMETERS}}
\centering
\begin{tabular}{cc}
\hline
\hline
\textbf{Parameters} & \textbf{Values} \\
\hline
Bandwidth  & 10 MHz \\
Carrier Frequency  & 8 GHz \\
Subcarrier spacing  &  15 KHz \\
Maximum Doppler shift ($\nu_{\text{max}}$) & 2 \\
Maximum delay spread ($\tau_{\text{max}}$) & 2 \\
Number of paths ($P$) & 4 \\
Spacing factor ($\zeta_\nu$) & 1 \\
\hline
\hline
\end{tabular}
\end{table}

\begin{figure}[htbp]
\label{fig3}
\centering
\subfloat[BER in terms of Tx IQI.]{
		\includegraphics[scale=0.58]{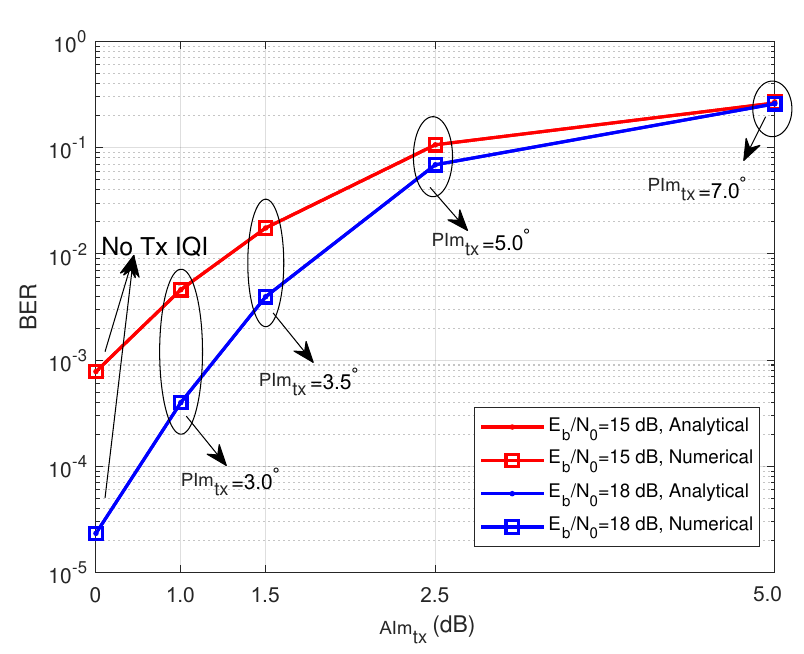}}
        
\subfloat[BER in terms of Rx IQI.]{
		\includegraphics[scale=0.58]{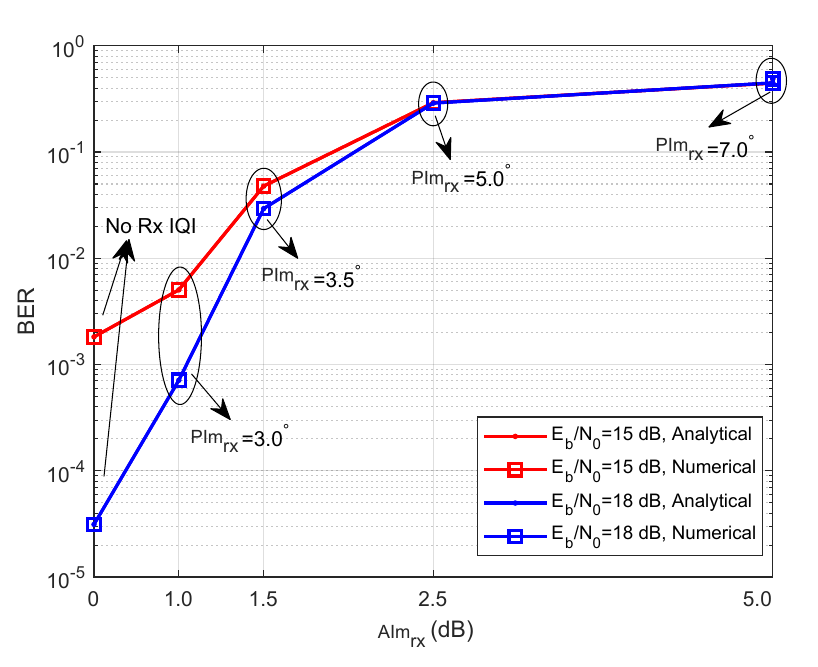}}
\caption{Analytical and numerical BER performance with ML detector in terms of: (a) Tx IQI, (b) Rx IQI.}
\end{figure}

In this simulation example, we evaluate the BER for QPSK modulated AFDM system by considering the joint effects of Tx and Rx IQI. Figs. 3(a) to 3(b) show the analytical and numerical results of BER as a function of IQI parameters at two fixed SNR values of 15 dB and 18 dB. It is observed that the numerical and analytical results are perfectly the same when the BER is evaluated in terms of the IQI parameters at fixed SNR value. In Fig. 3(a), the BER is depicted in terms of Tx IQI, i.e., the gain and phase mismatches pointed as $(\mathrm{AIm}_\text{tx},\mathrm{PIm}_\text{tx})$, while the Rx-IQI parameters have been fixed at the values of $(\mathrm{AIm}_\text{rx},\mathrm{PIm}_\text{rx})=(1.0\:\text{dB}, 3.0^\circ)$. It is seen that by increasing the gain and phase mismatches every step, the receiver performance degrades gradually. Obviously, in the values of $(\mathrm{AIm}_\text{tx},\mathrm{PIm}_\text{tx})=(2.5\:\text{dB}, 5^\circ)$ setting, the AFDM system becomes distorted. In Fig. 3(b), we evaluate the BER in terms of the gain and phase mismatches of Rx-IQI as $(\mathrm{AIm}_\text{rx},\mathrm{PIm}_\text{rx})$, while the Tx IQI parameters have been fixed at the values of $(\mathrm{AIm}_\text{tx},\mathrm{PIm}_\text{tx})=(1.0\:\text{dB}, 3.0^\circ)$. Apparently, compared with the Tx IQI, the Rx-IQI causes a more detrimental impact on the AFDM system. When the values of $(\mathrm{AIm}_\text{rx},\mathrm{PIm}_\text{rx})=(1.5\:\text{dB}, 3.5^\circ)$ sets, the distortion occurs. The reason is that the mirror interferences occur not only in the transmitted signals but also in effective channel response of the DAFT domain. 

\begin{figure}[!t]
\centering
\includegraphics[width=3.7in]{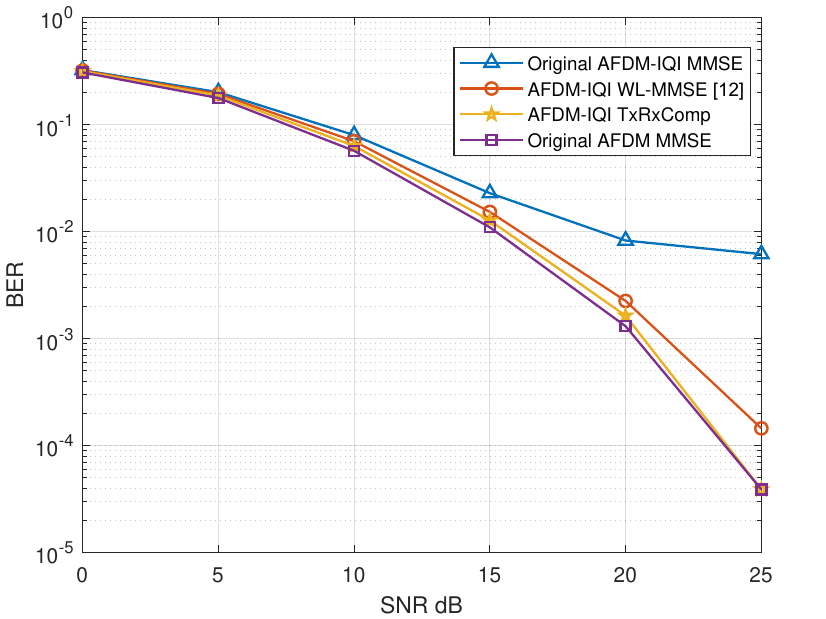}
\caption{The BER performance of the designed compensation method compared with baselines.}
\label{fig4}
\end{figure}

Furthermore, we evaluate the BER performance of the proposed compensation scheme in comparison with several benchmark methods, where $\mathrm{AIm}_\text{tx},\mathrm{AIm}_\text{rx}$ are set to 1 dB and 
$\mathrm{PIm}_\text{tx},\mathrm{PIm}_\text{rx}$ are $3^\circ$. The considered baselines include the MMSE detector for an ideal AFDM system, the MMSE detector for an AFDM system impaired by joint Tx and Rx IQI, and the WL-MMSE method in \cite{2025arXiv251210036}. It can be observed that the proposed compensation scheme paired with MMSE detection, almost completely recovers the performance loss caused by joint Tx and Rx IQI. By contrast, the WL-MMSE algorithm exploits only the covariance and pseudo-covariance of the improper noise induced by Rx IQI, and therefore cannot fully mitigate the resulting impairment. In addition, the computational complexity of the proposed compensation scheme combined with MMSE detection is $\mathcal{O}(N^3+2N)$, where the primary burden is the $\mathcal{O}(N^3)$ operations of the MMSE matrix inversion. In contrast, the computational complexity of WL-MMSE is $\mathcal{O}((2N)^3)$, since it jointly processes the received signal and its conjugate counterpart. At this complexity scale, the additional overhead introduced by the proposed compensation scheme is negligible.

\section{Conclusion}
In this paper, we first derive the input–output relationship of AFDM in the presence of joint Tx and Rx IQI, and further obtain analytical expressions for the PEP and ABEP in terms of the IQI parameters. To suppress the resulting distortion, we propose a cascaded compensation scheme, which first compensates for the Rx IQI in the time domain, converting the improper Gaussian noise back into AWGN, and then compensates for the Tx IQI based on the detected symbols. The proposed compensation scheme can be readily integrated with various AFDM detection methods without incurring additional computational complexity. Both analytical and simulation results demonstrate that joint Tx and Rx IQI causes severe performance degradation in AFDM systems, whereas, the proposed scheme effectively mitigates such impairments.

\bibliographystyle{IEEEtran}
\bibliography{references}

\end{document}